\documentclass[a4paper,11pt]{article}
\pdfoutput=1 
\usepackage{jinstpub} 
\makeatletter
\def\@fpheader{\relax} 
\makeatother
\usepackage{dcolumn}
\usepackage{upgreek}
\usepackage{lscape}
\usepackage{tabularx}
\usepackage{graphicx}
\usepackage{hyperref}
\usepackage[load-configurations=abbreviations,detect-all=true]{siunitx}
\usepackage[version=3]{mhchem}
\usepackage[mathlines]{lineno}
\usepackage{natbib}
\usepackage[inline,]{enumitem}

\input{defs.tex}

\title{\boldmath \TM CMOS direct charge sensing plane for neutrinoless double-beta decay search in high-pressure gaseous TPC}

\author[a]{Yuan Mei,}
\author[b]{Xiangming Sun,}
\author[a,b,c]{and Nu Xu}

\affiliation[a]{Lawrence Berkeley National Laboratory, Berkeley, California, USA}
\affiliation[b]{Central China Normal University, Wuhan, Hubei, China}
\affiliation[c]{Institute of Modern Physics, Chinese Academy of Sciences, Lanzhou, Gansu, China}

\emailAdd{ymei@lbl.gov}
\emailAdd{xmsun@phy.ccnu.edu.cn}

\abstract{We propose a novel charge sensing concept for high-pressure Time Projection Chamber (TPC) to search for Neutrinoless Double-Beta Decay (\znbb) with ton-scale isotope mass and beyond.  A meter-sized plane, tiled with an array of CMOS integrated sensors called \TM that directly collect charge without gas avalanche gain, is to be deployed into a high-pressure gaseous TPC with working gases containing suitable \znbb candidate isotopes such as \Xeots and \Seet.  The \TM sensor has an electronic noise $<\SI{30}{e^-}$ per pixel, which allows the detector to reach $<\SI{1}{\percent}$ FWHM energy resolution at the \znbb Q-value for both \Xeots and \SeetFs gases by measuring ionization charges alone.  The elimination of charge avalanche gain allows the direct sensing of slow-drifting ions, which enables the use of highly electronegative gas \SeFs in which free electrons do not exist.  It supports the swapping of working gases without hardware modification, which is a unique way to validate signals against radioactive backgrounds.  Since the sensor manufacturing and plane assembling could leverage unaltered industrial mass-production processes, stability, uniformity, scalability, and cost-effectiveness that are required for ton-scale experiments could all be reached.  The strengths of TPC such as 3D ionization tracking and decay daughter tagging are retained.  This development could lead to a competitive \znbb experiment at and above ton-scale.  The conceptual considerations, simulations, and initial prototyping are discussed.}

\begin{document}
\maketitle
\flushbottom

\section{Introduction}
\label{sec:intro}

Nuclear physics has played a very prominent role in the discovery of new neutrino physics---neutrino oscillations and hence the non-zero neutrino mass.  The field's next major goal is to uncover the nature of that mass.  Neutrinos, lacking charge or any other additively conserved quantum number, can support both Dirac and Majorana mass mechanisms.  In fact, the presence of both mechanisms allows us to simply explain the extreme lightness of neutrinos through the seesaw mechanism.  The Majorana mass, which implies that neutrino is its own antiparticle, would allow Neutrinoless Double-Beta Decay to occur.

Certain nuclei that are energetically forbidden to undergo single-beta decay could decay through a second-order weak process by simultaneously emitting two electrons (double-beta).  Normally two neutrinos are emitted in the process ($2\nu$).  If neutrino is its own antiparticle (Majorana), there exists a non-zero probability that no neutrino is emitted ($0\nu$) in the double-beta decay process (\znbb).  A positive experimental identification of \znbb will unambiguously prove the Majorana nature of neutrinos and provide a measure of the absolute neutrino mass.  \znbb is arguably the most sensitive experimental means to probe lepton number violation.

\subsection{Current status and future prospects of \znbb experiments}

The experimental signature of \znbb explored by most of the current generation experiments is a sharp peak at the Q-value, \Qbb, in the total beta energy spectrum.  Many techniques have been explored to measure such a signature.  The leading experiments, to name a few, include KamLAND-Zen\cite{Gando2016} and SNO+\cite{Andringa2015tza}, which load \znbb candidate isotopes into liquid scintillators and measure the energy spectrum through light output; Majorana\cite{Abgrall:2013rze,PhysRevLett.120.132502} and GERDA\cite{Agostini20161876,PhysRevLett.120.132503} (and henceforth the newly formed collaboration LEGEND\cite{Abgrall:2017syy}), which directly measure the ionization charge in germanium crystals; CUORE\cite{Artusa2015,PhysRevLett.120.132501}, which measures the temperature rise of \ce{TeO2} crystals caused by heat released from decay events; and EXO\cite{Albert2014,PhysRevLett.120.072701}/nEXO\cite{Pocar:2015ota}, which measure the ionization charge in a liquid xenon Time Projection Chamber.  Most of the aforementioned leading experiments have set a similar lower limit of \znbb half-life for various isotopes at \SI{E25}{year} level, with KamLAND-Zen setting the best limit for \Xeots to be $T_{1/2}^{\znbb}>\SI{1.07E26}{yr}$\cite{Gando2016}.

The current goal of the \znbb field is to push the sensitivity on the effective Majorana mass, \mbb, to below the allowed parameter space of inverted neutrino-mass ordering, or about \SI{10}{meV}.  It translates to a half-life limit $T_{1/2}^{\znbb}\gtrsim\SI{E28}{yr}$.  Such sensitivity requires an experiment with a ton of active isotope to run for several years, while achieving a background level of less than 0.1 counts per ton-year in the energy region of interest (ROI)\cite{Agostini:2017jim}.  Since in the presence of non-zero backgrounds the sensitivity scales as $T_{1/2}^{\znbb}\propto\sqrt{\frac{M\cdot t}{B\cdot\Delta E}}$, where $M$ is the total mass of candidate isotope, $t$ is the total observation time, $B$ is the background rate and $\Delta E$ is the energy resolution, experiments have to scale up to a large mass while simultaneously maintaining an excellent energy resolution and a very low radioactive background level.  While background and energy resolution are both of uttermost importance, the key challenge is to scale up the experiment to ton-scale and beyond, while not sacrificing either.  Current leading experiments, one way or another, face challenges in scaling up their respective technologies in the desirable fashion.


Among the current and planned experiments, the high-pressure gaseous TPC experiments, represented by NEXT\cite{Gomez-Cadenas2013} and \PXIII\cite{Chen2016}, although not currently leading the field, stand out for the reason that they have the best potential of fulfilling all the requirements for a next-generation experiment when scaled up.

\subsection{High-pressure gaseous Time Projection Chamber}

The gaseous Time Projection Chamber played an important historical role in the search of rare nuclear decays with a half-life exceeding $10^{20}\si{yr}$.  The very first direct observation of two-neutrino double-beta decay (\tnbb) was performed by placing a thin source containing tens of grams of \Seet in the center of a TPC\cite{Moe2014}.  The unique capability of the gaseous TPC to observe the double-beta ionization tracks to separate \tnbb events from backgrounds was critical to the success of the experiment.  In the \znbb era, the importance of the signal/background discrimination through $\beta$ tracking remains.  In addition, for \znbb, unlike the aforementioned TPC in which the source \Seet and the detector medium gas are separate, the detector medium and the \znbb candidate isotope have to be one in order to accommodate the large isotope quantity and to control potential background contamination.


It has recently been demonstrated that high-pressure gas such as xenon exhibits excellent intrinsic energy resolution (better than \SI{0.5}{\percent} FWHM) in the \znbb energy range\cite{Alvarez2013101} by ionization alone.  This energy resolution is not spoiled when the gas volume is monolithically increased.  When the xenon volume containing the \znbb candidate isotope \Xeots is instrumented as a TPC, it satisfies the detection medium-source unification and enables charge track imaging capabilities for discriminating double-beta signals against backgrounds.  TPCs are also easily scalable to large mass since the medium is monolithic and uniform.  It also opens doors for live detection of decay daughter such as single barium ion tagging\cite{Jones:2016qiq,PhysRevLett.120.132504}, which is yet another independent handle for double-beta to background discrimination.


The currently best-performing charge-readout technology to take advantage of both excellent energy resolution and charge tracking is electroluminescence, which is chosen by NEXT\cite{Gomez-Cadenas2013}.  While proven to retain excellent energy, electroluminescence has limitations in tracking and scalability.  The alternatives, such as MicroMegas (chosen by \PXIII in the first stage) and GEM, involve avalanche gas gain that severely deteriorates the energy resolution\cite{Alvarez201403,Alvarez201404}.  Simultaneous readout of many pixels is a challenge as well.

A pixelated charge readout plane without gas-electron avalanche is desirable.  The immediate challenge is to develop a sensor with very low noise so that even without electron multiplication, the energy resolution requirement could be met.  If such a device could be made, an array of these devices will be intrinsically stable and scalable to large sizes, hence eliminating the current shortcomings in instrumenting the high-pressure gaseous TPC.  This proposal addresses the challenges in realizing such a device and a readout with an array of these devices.

The proposed charge readout represents a clear path for upgrading gaseous TPC \znbb experiments.  At $\sim\SI{100}{kg}$ level, NEXT (xenon gas) predicts\cite{Martin-Albo:2015rhw} a background of \SI{10}{cts/ton.yr} with \SI{1}{\%} FWHM energy resolution.  Photo-sensors---PMTs and SiPM plane---contribute over half of the backgrounds.  By employing the \TM readout, which eliminates photo-sensors, the background index can be halved.  Many challenges in manufacturing, uniformity, and stability are also eliminated in the process, which allows the experiment to scale up to the desired \SI{1}{ton} active mass.  The readout is compatible with live barium tagging techniques\cite{Jones:2016qiq,PhysRevLett.120.132504}, which will allow the background index to approach the target \SI{0.1}{cts/ton.yr}.  Being able to detect drifting ions directly, the readout can instrument electronegative gas \SeFs.  Since the \znbb candidate isotope \Seet has a high $\Qbb\approx\SI{2995.5}{keV}$, where the external backgrounds are significantly lower than that around the \Qbb of \Xeots, a \SI{0.1}{cts/ton.yr} background index can be achieved as well\cite{Nygren:2018zsn}.  These requirements drive the detailed decisions towards a avalanche-free full CMOS readout that will achieve \SI{1}{\percent} FWHM energy resolution and a sub-leading background contribution.

\section{Approach and methods -- \TM charge readout plane}
\label{sec:TM}

We propose to realize a charge readout plane with a tiled array of CMOS charge sensors named \TM, which directly collect ionization charges without gas-electron multiplication.  Charge collection electrodes, front-end amplifiers, as well as data processing circuits, are integrated into every CMOS sensor placed directly at the site of charge measurement.  The readout scheme will simultaneously achieve the necessary low electronic noise to achieve a \SI{1}{\percent} FWHM energy resolution at \Qbb and sufficiently high spatial resolution for ionization charge tracking while satisfying the stringent radiopurity and scalability requirements.

By ionization alone, high-pressure gases exhibit excellent intrinsic energy resolutions of better than \SI{0.5}{\percent} FWHM\cite{Alvarez2013101} in the vicinity of the \Xeots $\Qbb\approx\SI{2458}{keV}$ and \SeetFs $\Qbb\approx\SI{2995.5}{keV}$.  On average each \znbb event liberates about \num{1e5} ionization electrons (or ions.  Hereinafter we use electrons to refer to charges in general).  If the total noise (fluctuation) contribution of the readout is suppressed below $370$ electrons ($\sigma$) on an event-by-event basis, the \SI{1}{\percent} FWHM energy resolution could be reached.  A pixelated charge readout plane that directly measures charge without gas-electron avalanche could achieve this goal.  Without gas gain, the charge sensing electronics must have exceedingly low internal noise to achieve sufficient signal-to-noise ratio.  The front-end must be placed very close to the charge collection site, and the signal must be digitized as early as possible and transmitted digitally to minimize cross-talk and interference.  All technical requirements point towards a charge-collecting analog-digital mixed integrated circuitry.

\begin{figure}[!htb]
  \centering
  \begin{minipage}[t]{0.25\linewidth}
    (a)\\[6ex]
    \includegraphics[width=\linewidth]{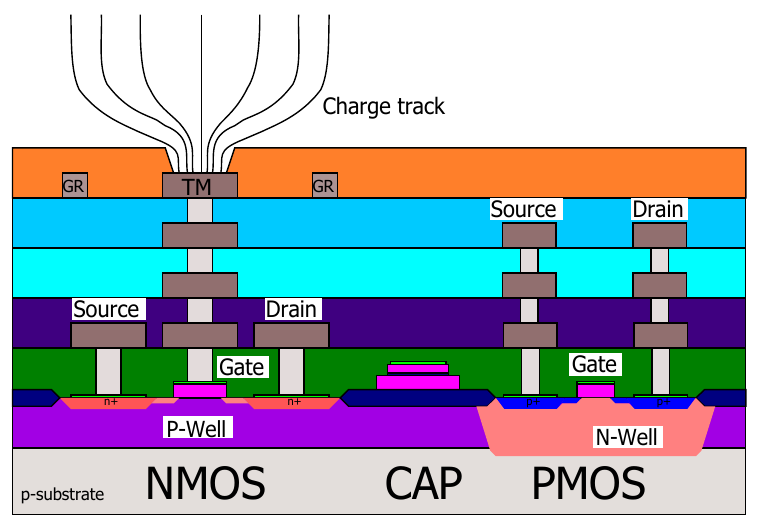}
  \end{minipage}\hspace{0.5em}%
  \begin{minipage}[t]{0.30\linewidth}
    (b)\\[2ex]
    \includegraphics[width=\linewidth]{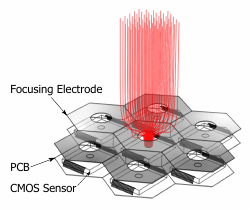}
  \end{minipage}\hspace{0.5em}%
  \begin{minipage}[t]{0.35\linewidth}
    (c)\\
    \includegraphics[width=\linewidth]{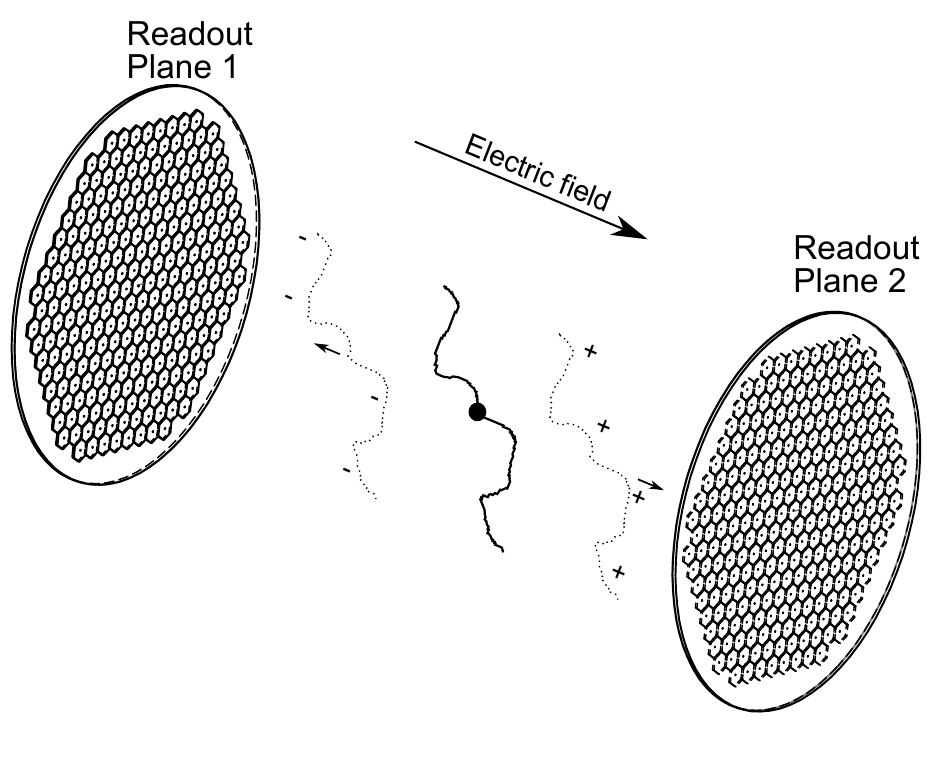}
  \end{minipage}
  \caption{\TM sensor and array for high-pressure gas TPC.  (a) Cross section view of a \TM CMOS sensor.  (b) \TM sensors tiled in a hexagonal pattern to form a charge readout plane without gas gain.  Electrode size is $\sim\SI{1}{mm}$ and the pitch is $5\sim\SI{10}{mm}$.  Focusing electrode with hole-pattern matching the sensor array is placed above the array to improve the charge collection efficiency.  Red lines are simulated charge drifting trajectories.  (c) Conceptual design of a TPC using \TM charge readout plane.}
  \label{fig:tmTPC}
\end{figure}

Fig.~\ref{fig:tmTPC}(a) illustrates the cross section view of a proposed CMOS charge sensor \TM.  A metal patch on the topmost layer of the sensor is exposed to directly collect charge.  The electrode is immediately connected to a transistor circuitry underneath the electrode to convert charge into voltage signal, which is then processed and digitized by additional circuitry implemented in the same sensor.  A conceptual design of a charge readout plane with many \TM sensors arranged in hexagonal pattern is shown in Fig.~\ref{fig:tmTPC}(b).  A focusing electrode with perforated round-hole pattern matching the array is placed above the plane with openings aligned with each electrode concentrically.  The focusing structure ensures all charges eventually land on the \TM electrode for maximum charge collection efficiency.

The advantage of such a charge readout scheme, in addition to the competitive energy resolution of $<\SI{1}{\percent}$ FWHM, is the stability and scalability.  The elimination of gas avalanche gain ensures long-term stable operation of the detector.  The industrial semiconductor production and assembly procedure guarantee uniformity, which is critical to the production of meter-sized planes (Fig.~\ref{fig:tmTPC}(c)).  Since no gas gain is necessary, both drifting ions and electrons could be detected.  Silicon CMOS process and circuit board materials and assembly procedures are radiopure as well.  The reasoning for the design choices is described in the following sections.

\subsection{Design considerations}

The success of the proposed charge readout scheme relies on the realization of the following essential elements:
\begin{enumerate}[noitemsep,topsep=-0.5ex]
\item Deciding on an optimized pitch spacing between pixels (sensors), trading off among diffusion, energy resolution, spatial resolution, etc.
\item A CMOS sensor that directly collects and measures charge.  Collected charge is converted to a voltage signal by means of a Charge Sensitive pre-Amplifier (CSA) and then digitized on-site in the sensor.  The Equivalent Noise Charge (ENC) should be $<\SI{30}{e^-}$.
\item A charge focusing structure above the sensor that concentrates and guides drifting charge to land on the collection electrode of the CMOS sensor.  Charge loss in the process should be near zero.
\item Tiling about \num{1e5} sensors with optimized spacing ($5\sim\SI{10}{mm}$) on a meter-sized plane and routing all signals off the plane.  The greatest challenge is the signal path management.  Sensors should contain data processing and transmission units and are interconnected to form a data communication network.  In such a configuration sensors are predominantly interconnected with local neighbors so that the number of signal lines coming off the plane becomes manageable.  Fault tolerance should be built into the network so that failed sensors will not disable a large section of the network.
\item Construction using radiopure material.  CMOS chips are known to be low in radioactive contamination\cite{Leonard:2007uv}, and the total mass used in the experiment is also small (\SI{1}{kg} or less).  A select class of radiopure Printed Circuit Board (PCB) materials exist and have been in use in the field\cite{Albert2017}.  The employment of designs and assembly procedures that are compatible with the clean materials is required.
\end{enumerate}

\subsection{Pixel pitch}\label{sec:pp}

The decision on the pitch size between pixels is a result of the trade-off among achievable electronic noise of the CSA, charge diffusion in the gas, and background discrimination efficiency.  The leading constraint is to achieve a \SI{1}{\percent} or better FWHM energy resolution.

A feature of the CMOS pixel plane is that, within limits, the electronic noise of a single pixel is independent of the pixel pitch (or pixel size).  Given a circuitry, the electronic noise of the CSA is largely affected only by the input capacitance hence the \TM electrode size and shape.  As soon as the \TM electrode is determined, the electronic noise of the pixel is fixed.  However, the pitch between pixels could still vary provided the pitch does not become too large so that the electrostatic focusing could still collect all the charges and that there are no other undesirable effects.

\begin{figure}[!htb]
  \centering
    \includegraphics[]{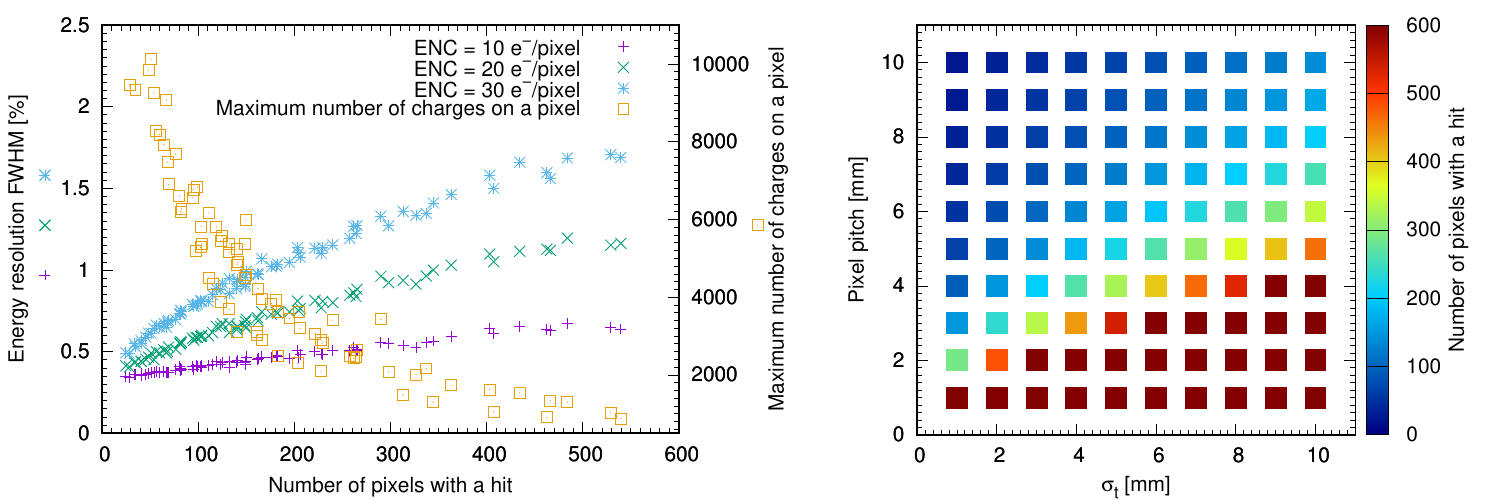}
  \caption{Energy resolution influenced by number of pixels that receive a non-zero charge signal (hit).  Left: Energy resolution as a function of total number of pixels that see signal.  Maximum number of charges on a single pixel as a function of the total number of pixels that see signal (yellow square) is overlaid.  Right: total number of pixels that see signal as a function of pixel pitch and transverse diffusion $\sigma_t$.}
  \label{fig:dpnEr1}
\end{figure}

A CSA with the noise performance (ENC) in the tens of electrons is achievable (see Sec.~\ref{sec:cssr}).  Given this number, the required energy resolution sets the upper limit of the total number of pixels ($N$) that acquire charge for each event.  The total electronic noise could be estimated as ENC$\times\sqrt{N}$.  A more elaborate simulation that includes the intrinsic fluctuation of high-pressure xenon gas shows in Fig.~\ref{fig:dpnEr1}(left) the dependence of energy resolution on $N$ for different ENC/pixel levels.  For $\text{ENC}=\SI{30}{e^-}$ per pixel, to reach a \SI{1}{\percent} FWHM energy resolution, at most $\sim150$ pixels are allowed to see a part of the track.

\begin{figure}[!htb]
  \centering
    \includegraphics[]{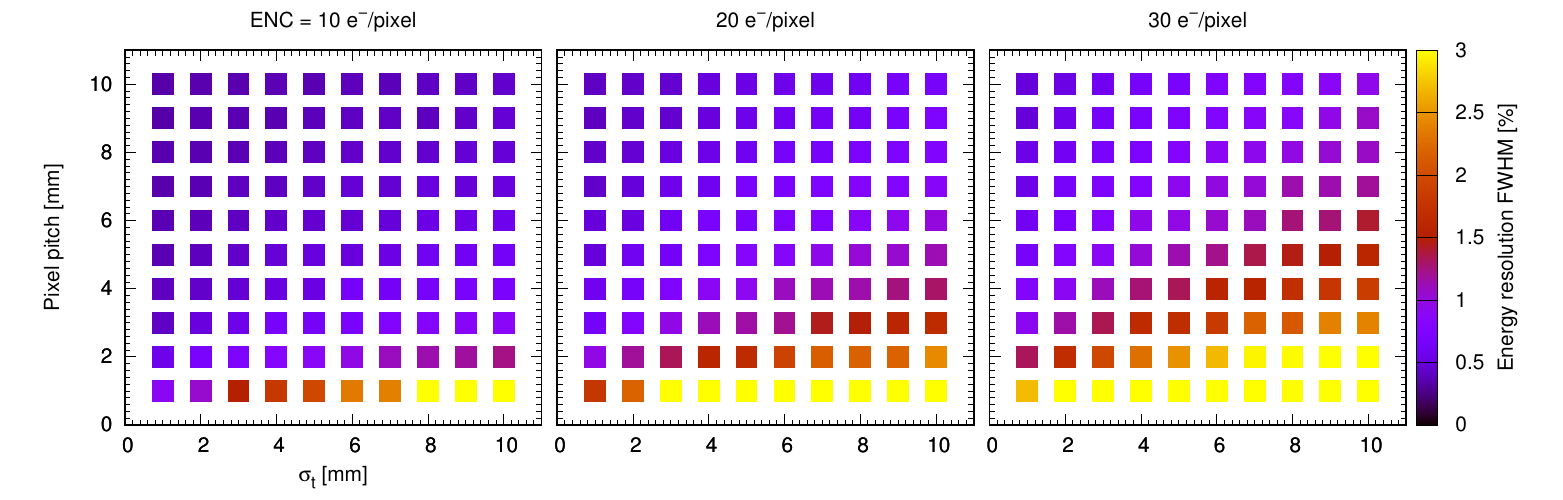}
  \caption{Impact of diffusion, pixel pitch and pixel noise on the energy resolution.  $x$-axis shows the transverse diffusion $\sigma_t$.  $y$-axis shows the pixel pitch.  The three sub-plots correspond to three ENC/pixel levels at 10, 20, and \SI{30}{e^-} respectively.}
  \label{fig:dpnEr}
\end{figure}

The constraint on the maximum total number of pixels ($N$) that acquire charge could be satisfied by balancing the pixel pitch and the transverse diffusion of charge $\sigma_t$.  It is best illustrated in Fig.~\ref{fig:dpnEr1}(right), which shows the maximum total number of pixels with a signal as a function of pixel pitch and $\sigma_t$.  For a desired maximum total number of pixels with a signal, multiple pairs of (pitch, $\sigma_t$) exist.  For \SI{1}{m} drift in pure xenon, the transverse diffusion in a \SI{500}{V/cm} drift field is $\sigma_{xy}\approx\SI{9}{mm}$\cite{Nygren2009337}.  A pixel pitch of $\sim\SI{8}{mm}$ would correspond to $\sim150$ pixel hits, which satisfies the aforementioned goal of \SI{1}{\percent} FWHM energy resolution.  The mapping from (pitch, $\sigma_t$) directly to energy resolution is shown in Fig.~\ref{fig:dpnEr}.  Such pitch size is within the limits imposed by the electrostatic focusing requirements as well.  I will use this value as a baseline design parameter.

Gas additives could lower the diffusion coefficient by a factor of up to 20\cite{GonzalezDiaz20158,Irastorza2016}.  Drifting ions could achieve a similar reduction of diffusion.  With smaller diffusion, the pixel pitch could be reduced accordingly.  The lower-bound, however, is set by practicality and cost as to how many sensors can be produced and how densely they can be packed, as well as the diminishing gain on the background discrimination from pattern recognition (see Sec.~\ref{sec:ctpr}).


The pixel pitch also sets the dynamic range that the CSA and the readout electronics must cover.  Fig.~\ref{fig:dpnEr1}(left) shows a simulated maximum amount of charge on a single pixel as a function of the number of pixels with a hit.  With the maximum number of pixels being $150$, every pixel should cover a signal range up to \num{5e3} electrons.

\subsection{Charge sensing and signal recovery}\label{sec:cssr}

\begin{figure}[!htb]
  \centering
  \begin{minipage}[t]{0.25\linewidth}
    (a)\\
    \includegraphics[width=\linewidth]{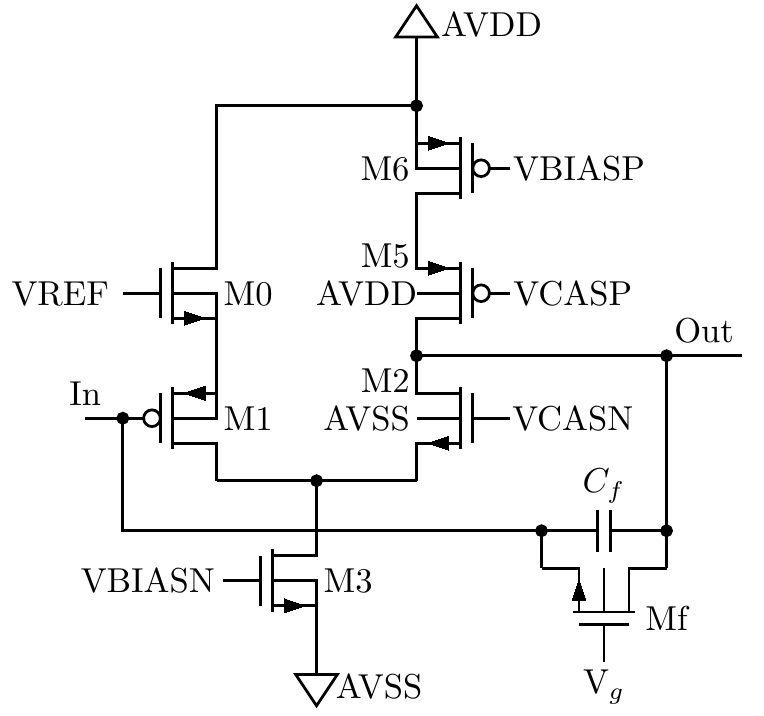}
  \end{minipage}\hspace{1pt}%
  \begin{minipage}[t]{0.37\linewidth}
    (b)\\
    \includegraphics[width=\linewidth]{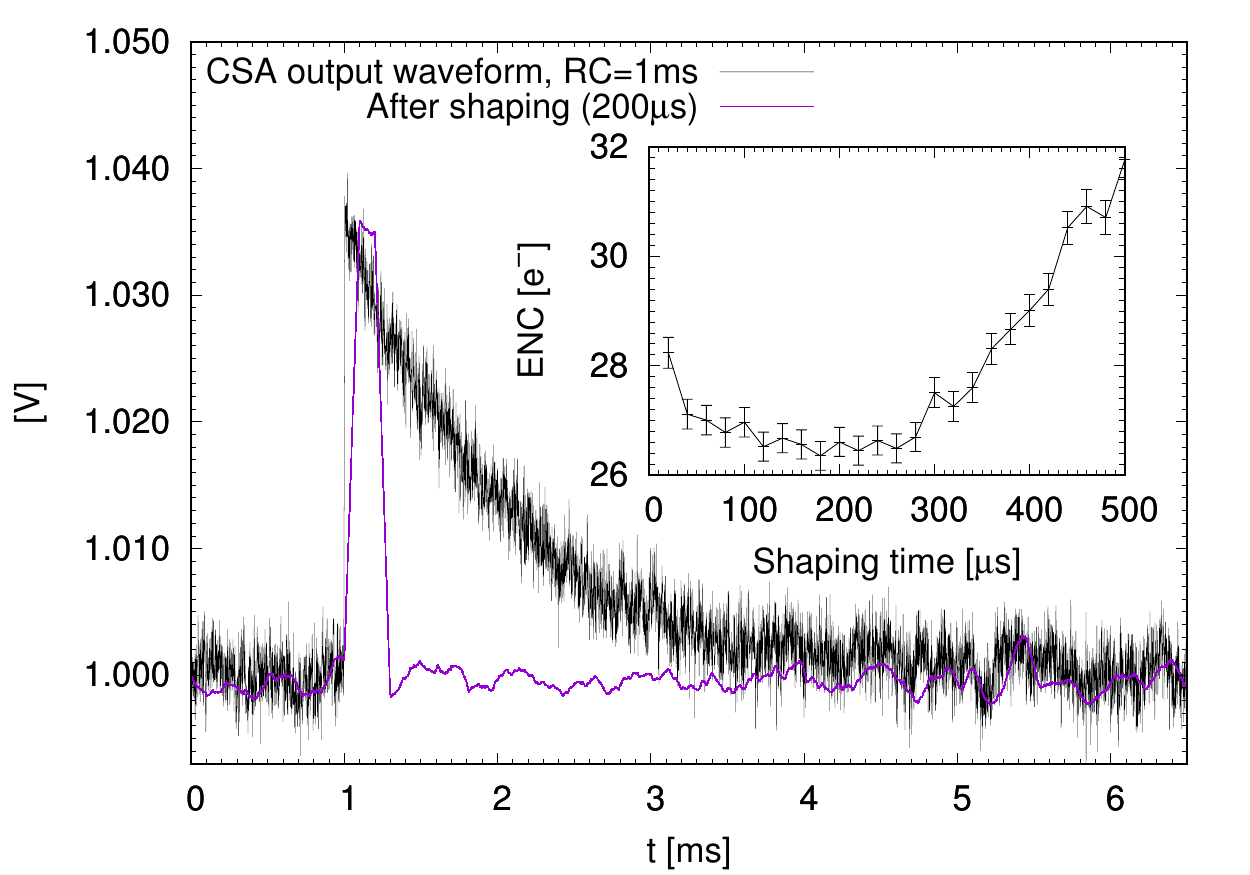}
  \end{minipage}\hspace{1pt}%
  \begin{minipage}[t]{0.37\linewidth}
    (c)\\
    \includegraphics[width=\linewidth]{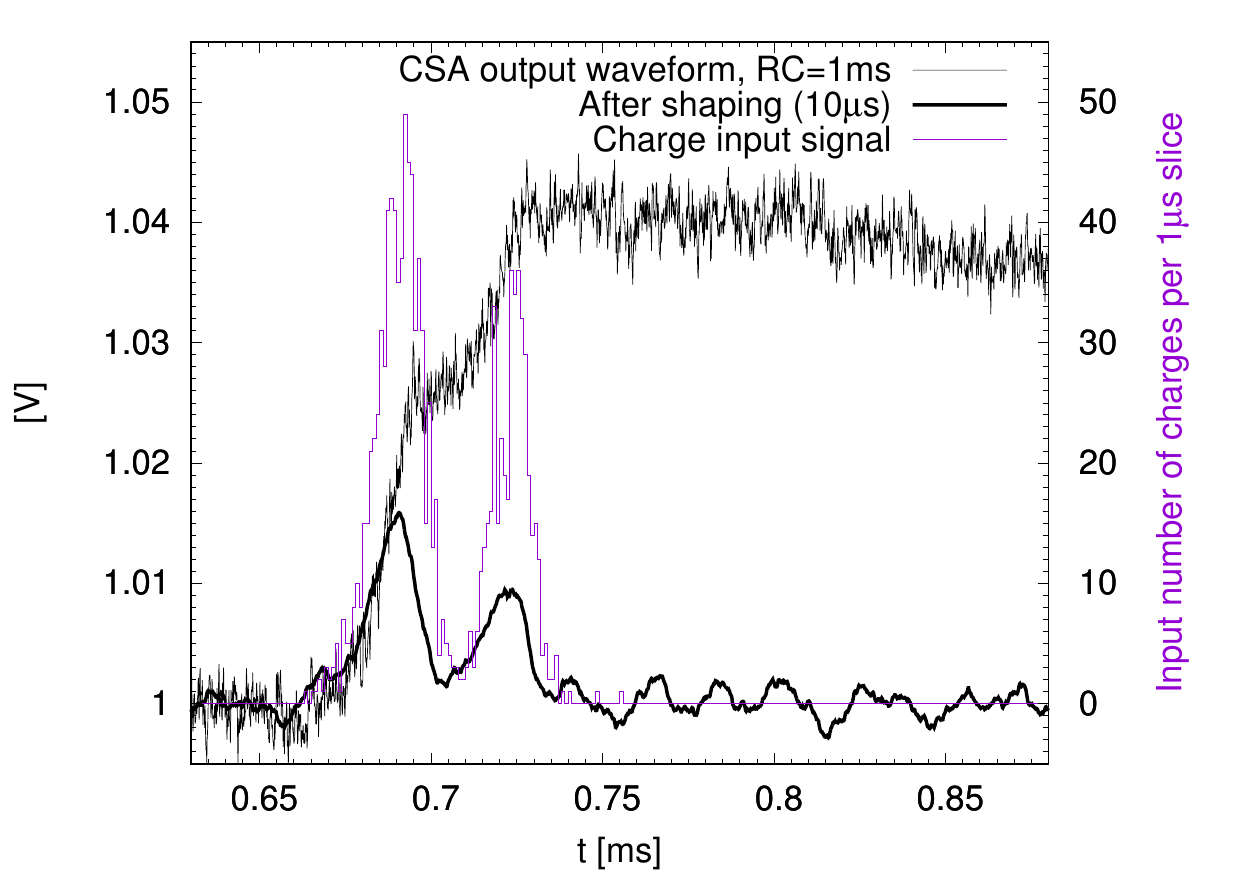}
  \end{minipage}\hspace{1pt}%
  \caption{Charge Sensitive pre-Amplifier (CSA).  (a) Schematic of a CSA implemented in the prototype sensor.  \sym{AVDD} and \sym{AVSS} are power and ground respectively.  All labeled voltages are provided by dedicated biasing circuitry (not shown).  (b) Expected output of CSA responding to a charge impulse input of \SI{1000}{e^-}.  Purple line shows the resulting waveform from a trapezoidal pulse shaper\cite{Jordanov1994}.  Inset shows the Equivalent Noise Charge (ENC) dependence on shaping time.  (c) A distributed charge input (purple line), the CSA response (thin black line), and the recovered waveform after a shaper (thick black line).}
  \label{fig:CSA}
\end{figure}

The charge signal collected on the \TM electrode is directly DC-coupled into a Charge Sensitive pre-Amplifier (CSA).  The structure of a simple CSA in the prototype sensor is shown in Fig.~\ref{fig:CSA}(a).  The CMOS process allows the manufacturing of a feed-back capacitor with a small but well controlled capacitance \Cf around \SI{5}{fF}.  With an amplification stage of sufficiently high gain, the input charge is entirely transferred onto \Cf; therefore, the input charge $Q$ to output voltage pulse height $V_o$ conversion gain is predominantly set by \Cf with the relation $V_o\approx Q/\Cf$.  This particular CSA has such a conversion gain of about \SI{35}{mV}/\SI{1000}{e^-}.  A reset transistor \sym{Mf} parallel to \Cf provides a channel for charge release and baseline restoration.  The gate voltage of \sym{Mf} is precisely tuned such that \sym{Mf} is close to be completely shut off so that its equivalent resistance $R$ is large.  Combined with \Cf, the $RC$ constant could be tuned by varying the gate voltage, and its value could practically extend into many milliseconds.  A CMOS CSA working in such a regime has been validated in one of our earlier sensors\cite{TopmetalII-2016}, which reached a $<\SI{15}{e^-}$ noise.


The challenge for \znbb optimized sensor is that the size of the charge collection electrode, hence its capacitance with respect to ground, \Cd, which is also the input capacitance to the CSA, is large.  For an electrode of \SI{1}{mm} in diameter, its $\Cd\approx\SI{5}{pF}$.  The noise of the CSA is adversely affected by the large input capacitance.  By carefully tuning the parameters of the circuitry in Fig.~\ref{fig:CSA}(a), the design goal of $\text{ENC}<\SI{30}{e^-}$ can be reached.  Fig.~\ref{fig:CSA}(b) shows a simulation of the CSA output responding to a test pulse.  The lowest noise is achieved at a software-defined shaping time of about \SI{180}{\micro s} using a digital trapezoidal filter\cite{Jordanov1994}, which is appropriate for total charge (energy) measurement.  There are incentives to make the electrode larger since the charge collection efficiency would improve.  The positive correlation between electrode size and noise sets an upper limit on the size to about \SI{1}{mm}.  Therefore, I decide on the \SI{1}{mm} electrode size while relying on the electrostatic focusing to achieve the desired charge collection.

The rise-time of the CSA also meets the experimental requirements.  Fig.~\ref{fig:CSA}(c) shows that the arrival of two consecutive charge pulses within the realistic time-scale of a single \znbb event could be reconstructed by processing the output of the CSA.  Such reconstruction determines the spatial resolution in the $z$ direction, which is estimated to be $<\SI{1}{mm}$ considering the CSA alone.  The simulated signal is obtained from the combination of both double-beta energy deposition in gas using Geant4 and the ionization charge drifting in gas through the electrostatic structure discussed in Sec.~\ref{sec:elecfoc}.

It is worth noting that when high spatial resolution in $z$ is desired, a narrower shaper has to be used.  A narrower shaper results in worse energy resolution; however, since the entire waveform is recorded and the shaper is applied afterwards in software, the spatial resolution in $z$ and the total energy resolution requirements do not contradict each other.  A wide shaper is used to extract total energy with optimal resolution while a narrow shaper is used to determine the charge distribution along the $z$ axis.

\subsection{Signal digitization and data transmission}

Due to the stringent noise requirement, the analog signal---the output of the CSA---must be digitized immediately inside of the sensor.  An in-chip Analog-to-Digital Converter (ADC) is required.  The ADC should have a noise floor well below the noise of the CSA, which is \SI{30}{e^-} or equivalently about \SI{1}{mV}.  It should also have a large enough number of bits (dynamic range) to cover the possible range of charge input (\SI{5e3}{e^-} or equivalently \SI{175}{mV}).

In addition, since the sensors are densely packed on the plane, the number of available traces for routing signal out of the plane is limited.  Beyond certain plane size (total number of sensors), routing every signal from all sensors out becomes impractical.  Digitized data must be communicated through inter-sensor network; therefore, circuitry that handles data processing and communication must be integrated in the sensor.

\subsection{Electrostatic focusing}
\label{sec:elecfoc}

As shown in Fig.~\ref{fig:tmTPC}(b), a sheet electrode with perforated round-hole pattern matching the array is placed above the plane with openings aligned with each electrode on the CMOS sensor concentrically.  The electrode is electrostatically biased to a high voltage such that the electric field between the sheet electrode and the sensors is much higher than that in the drift region.  Charges drifting towards the plane will congregate towards the center of an opening then pass through and land on the sensor electrode (focusing).  Red curves in Fig.~\ref{fig:tmTPC}(b) are simulated trajectories of the drifting charges being focused.

For \SI{8}{mm} pixel pitch and \SI{1}{mm} electrode size, a factor of 64 increase of the electric field in the focusing region with respect to the drifting region is required.  Assuming a \SI{500}{V/cm} field strength in the drifting region, the field strength in the focusing region would be \SI{32}{kV/cm}.  In \SI{10}{bar} xenon gas, such an electric field is slightly above the onset of electroluminescence but is safely below electron-gas avalanche threshold\cite{Nygren2009337}.  Moderate number of electroluminescence photons generated will not pose a problem since the sensors are insensitive to light.  Moreover, the main drifting field could be reduced to as low as \SI{200}{V/cm} without any adverse consequences, which would further moderate the potential issue.

The simulation shows that the aforementioned field configurations achieve full charge collection, i.e.\ all charges that come from the drift region will land on the sensor electrode (Fig.~\ref{fig:tmTPC}(b)).  In certain local regions, the field strength deviates from the idealized arguments above.  The actual configuration that achieves the best focusing is found by sweeping the parameters that include the electrode geometry, the distance between the focusing electrode and the sensor, and the high-voltage potential.

In case lowering the drift field is undesirable, an umbrella-shaped electrode could be chemically grown on top of the \TM electrode.  The electrode will extend the \TM vertically (the stem) by about a \si{mm} first, then horizontally (the canopy).  Such a structure will not add additional capacitance hence has minimal impact on the noise but will add significant area for charge collection, which will allow a lower field in the focusing region to achieve the same focusing effect.

For the strongly electronegative gas \SeFs, the electric field breakdown threshold is much higher than that of xenon.  Therefore, the field strength can be increased by a large factor to ensure the full charge collection and to increase the charge drift velocity.  Refer to \cite{Nygren:2018zsn} for further details.

\subsection{Charge track pattern for background discrimination}
\label{sec:ctpr}

\begin{figure}[!htb]
  \centering
    \includegraphics[]{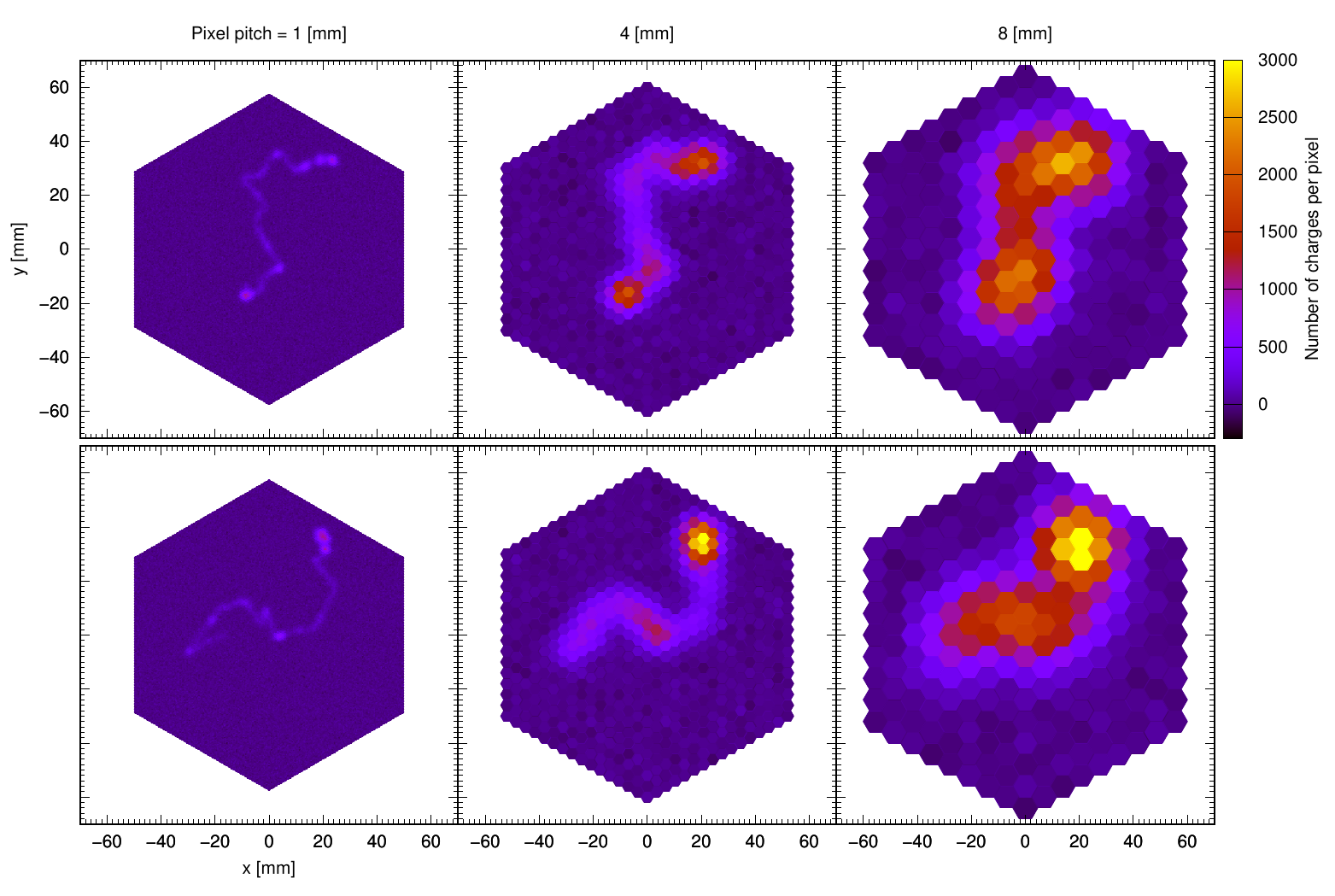}
  \caption{Charge track comparison of a double beta-decay (\dbd) event (top row) and a background event (bottom row) under various diffusion and pixelation conditions.  Diffusion ($\sigma_t$) is set to equal pixel pitch.  Images show the $xy$ projection of 3D tracks.  Pure xenon corresponds to the \SI{8}{mm} pitch (right most) case.  Gas additives could reduce the diffusion.  \SI{30}{e^-} noise per pixel is added.  The two-blob feature of the \dbd event and one-blob for background are clearly visible regardless of the pixel pitch size.}
  \label{fig:Ptn0nBBvBeta}
\end{figure}

A feature that is unique to high-pressure gaseous TPC is that the 3D geometry of ionization charge tracks is extended and is well measured along with the spatial distribution of charge density.  Fig.~\ref{fig:Ptn0nBBvBeta} shows the projection of a representative double-beta decay (\dbd) event and a background of identical total energy in high-pressure xenon TPC.  As an energetic electron traverses the high-pressure gas, it loses energy and leaves a track of ionization along its path.  As the electron approaches the end of its track, the ionization density rises right before its full-stop (Bragg peak).  Therefore, every fully contained energetic track in the gas volume has a ``blob'' of high ionization density.  Since a \dbd event emits two energy electrons, its track has two blobs at the two ends.  This feature sets the \dbd event apart from backgrounds, which predominantly has only one blob at one end of the track.  Due to multiple Coulomb scattering, $\sim\si{MeV}$ tracks are not straight in a dense gas; however, the feature of blobs persists.

\begin{figure}[!htb]
  \centering
  \includegraphics[width=0.6\linewidth]{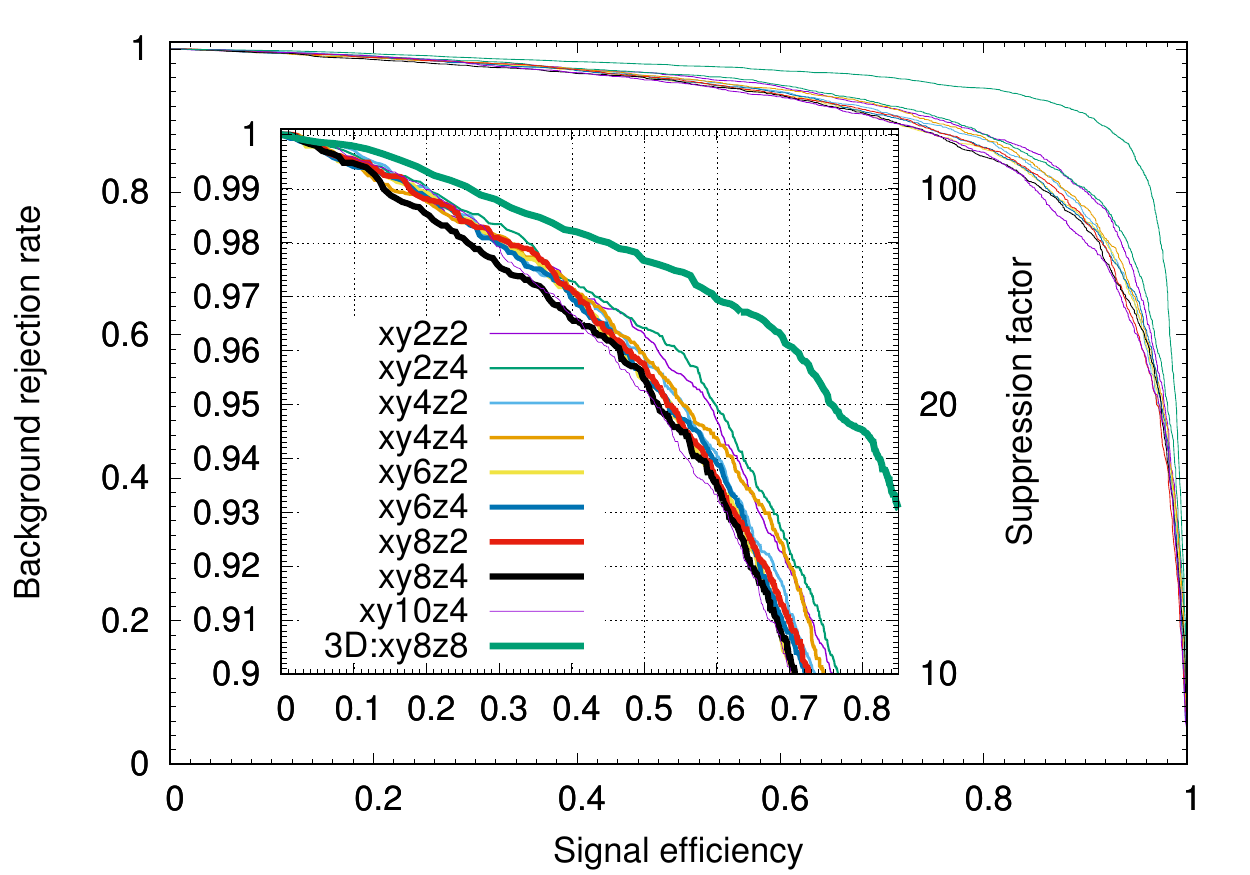}
  \caption{Receiver Operating Characteristic (ROC) curves of pattern recognition using various pixel (voxel) sizes.  ROC curves characterize the interplay between signal efficiency (acceptance) and background rejection rate.  Curves labeled without ``3D'' are results from 2D projection images of charge tracks used as input to the neural network.  From the figure the change in voxel size (labeled in the legend are \si{mm} sizes in $x,y$ and $z$ dimensions) has only moderate effects on the accuracy of signal-to-background discrimination.  However, going from 2D projections to a full 3D recognition shows a large improvement in background rejection rate.}
  \label{fig:PtnROC}
\end{figure}

Since the TPC could measure the three-dimensional shape of tracks and ionization density, pattern recognition methods have been developed specifically for high-pressure gas TPC to discriminate \dbd events against backgrounds (see for example \cite{Cebrian2013,Ferrario2016}).  Machine learning algorithms employing Deep Neural Networks (DNNs) have been applied to this problem as well\cite{Renner2016}.  We independently developed two DNNs that operate on simulated \dbd and background events to study the impact of the pixel size on the effectiveness of signal-background discrimination.  Firstly, we constructed an image processing DNN based on AlexNet\cite{Krizhevsky2012}.  The network takes the 2D projections of tracks onto $xy$, $yz$, and $zx$ planes as the input and classifies events as signal or background with an estimated probability.  The resulting relation between signal efficiency (acceptance) and background rejection rate is shown as the Receiver Operating Characteristic (ROC) curves in Fig.~\ref{fig:PtnROC}.  Results for various pixel sizes are overlaid in the figure.  In the second attempt, we trained a DNN based on ResNET\cite{He2015} that operates directly on the 3D distribution of charge tracks as input.  We conclude that given a method (2D or 3D), the pixel size has only minor effects on the accuracy of discrimination, but upgrading from 2D to 3D drastically improves the accuracy.

The above conclusion supports the choice of a somewhat large pixel pitch of \SI{8}{mm} that optimizes the energy resolution.  The exact trade-off point on the ROC curve should be determined as part of a global optimization of the data analysis.  As an example, with \SI{8}{mm} pixel size, if \SI{80}{\percent} signal acceptance is desired, a $\sim20:1$ two-electron signal to single-electron background rejection ratio can be achieved.  This ratio is the extra background suppression factor after all other cuts, including energy ROI.

In the strongly electronegative gas \SeFs, only ion drifting is possible and ions have less diffusion, therefore, better pattern based background discrimination could be achieved\cite{Nygren:2018zsn}.

\subsection{Sensor network}\label{sec:sn}

On a large plane there will be $\sim\num{1e5}$ sensors.  It will become practically impossible to route signals directly from every sensor to the edge of the plane for readout.  An obvious solution is to establish local or regional connections between nearby sensors to form a sensor network.  Each sensor in the network not only generates and transmits its own data but also relays data from other sensors.  A large selection of mature network topologies have been well studied and employed in industry\cite{Papamichael2012}.  Since \znbb is a rare event search experiment, the overall data rate is very low, and the data readout ``dead time'' is tolerable.  It is sufficient to implement a small data buffer in every sensor to store an event worth of data, then connect sensors using a segmented first-in-first-out (FIFO) chain structure.  The data will be shifted out in a serial fashion upon a global trigger decision.  This readout scheme implies that the data collection and transmission do not happen simultaneously.  It has an added advantage that there will be no activity on transmission signal lines during data collection so that the potential interference from the digital data transmission back to the very sensitive charge measurement circuitry is minimized.

An important aspect in this implementation is fault tolerance, namely a single failed sensor in the chain shall not disable the entire readout chain.  It could be achieved by incorporating configurable signal routing in the sensor.  The details of these features and options will be investigated during the IC design phase of the project.

\subsection{Plane assembly and material selection}\label{sec:pams}

Despite the large number of sensors, for \SI{8}{mm} pitch size, conventional wire-bonding is still the most reliable and cost-effective method for assembling sensors onto a PCB.  The PCB with multiple copper layers should be made with radiopure substrate materials such as Kapton.  The PCB shall be supported by a large copper plate for both mechanical stability and cooling.  Sensors should be placed onto the plane under the guidance of an optical alignment apparatus then wire-bonded using automated machines.  The perforated focusing electrode will be produced by photo-etching a thin copper sheet.  It is then stretched, mechanically aligned, and fixed above the plane.

A potential caveat is the external components such as resistors and capacitors which are sometimes required for the sensors to function properly.  For example, it is common practice to place decoupling capacitors next to CMOS sensors to condition the power supply.  These passive components together with the solder materials are known to be radioactive.  A solution is to absorb such components into the sensor.  It is natural to do so since new sensors will be designed and the sensor area could accommodate additional modules.  This option will be explored as part of the project.


\section{\TM sensor development and prototyping}

Prior to this proposal that is specific for \znbb, the authors developed the generic \TM CMOS direct charge sensor concept.  \TM is a series of charge sensors produced using the industrial standard Complementary Metal-Oxide-Semiconductor (CMOS) Integrated Circuit (IC) process.  Each sensor has exposed metal nodes for charge collection on its topmost layer, and integrates charge sensitive amplifier and signal processing circuitry right at the site of charge collection (Fig.~\ref{fig:tmTPC}(a)), hence achieving the optimal noise performance.

\begin{figure}[!htb]
  \centering
    \includegraphics[]{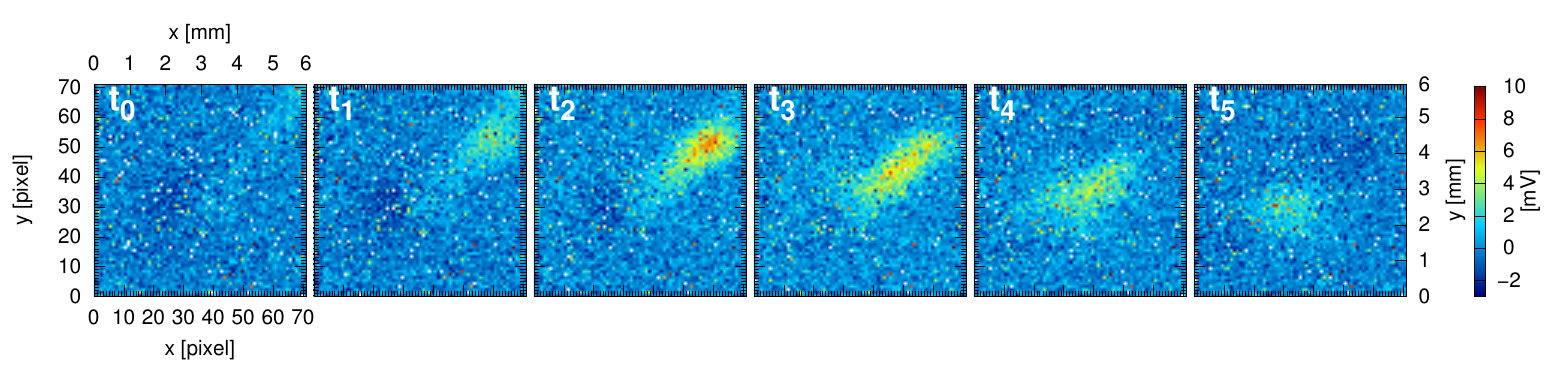}
  \caption{Single alpha particle ionization track, drifting through ambient air in an electric field, imaged by \TMIIm sensor.  Time interval between adjacent frames is about \SI{3.3}{ms}\cite{TopmetalII-2016}.}
  \label{fig:TMIImAlphaImg}
\end{figure}

Initially, \TM was conceived for imaging charge clouds in a TPC with high spatial resolution, therefore in its first two versions\cite{TopmetalI2014,TopmetalII-2016}, a high density pixel array ($\sim\SI{80}{\micro m}$ pitch size between pixels) was implemented in each CMOS sensor.  The second generation prototype, \TMIIm, demonstrated a noise level of less than \SI{15}{e^-} per pixel\cite{TopmetalII-2016}.  It also demonstrated that using the standard CMOS process without additional surface treatment, charges in gas can land on the sensor and be directly imaged.  Fig.~\ref{fig:TMIImAlphaImg} shows an ionization track in ambient air, which is liberated by one alpha particle from an \Amtfo source, drifting through air and imaged by a \TMIIm sensor.  It is worth noting that not only electrons, but also ions, are detected, which supports the proposed direct detection of ions in \SeFs.

Building on the past successful experiences with \TM for other applications, we have designed and produced the third generation sensor called \TMS, as a pilot device that is specifically optimized for \znbb in gaseous TPC.

\subsection{\TMS\ -- first sensor optimized for \znbb}

\begin{figure}[!htb]
  \centering
  \begin{minipage}[t]{0.75\linewidth}
    (a)\\
    \includegraphics[width=\linewidth]{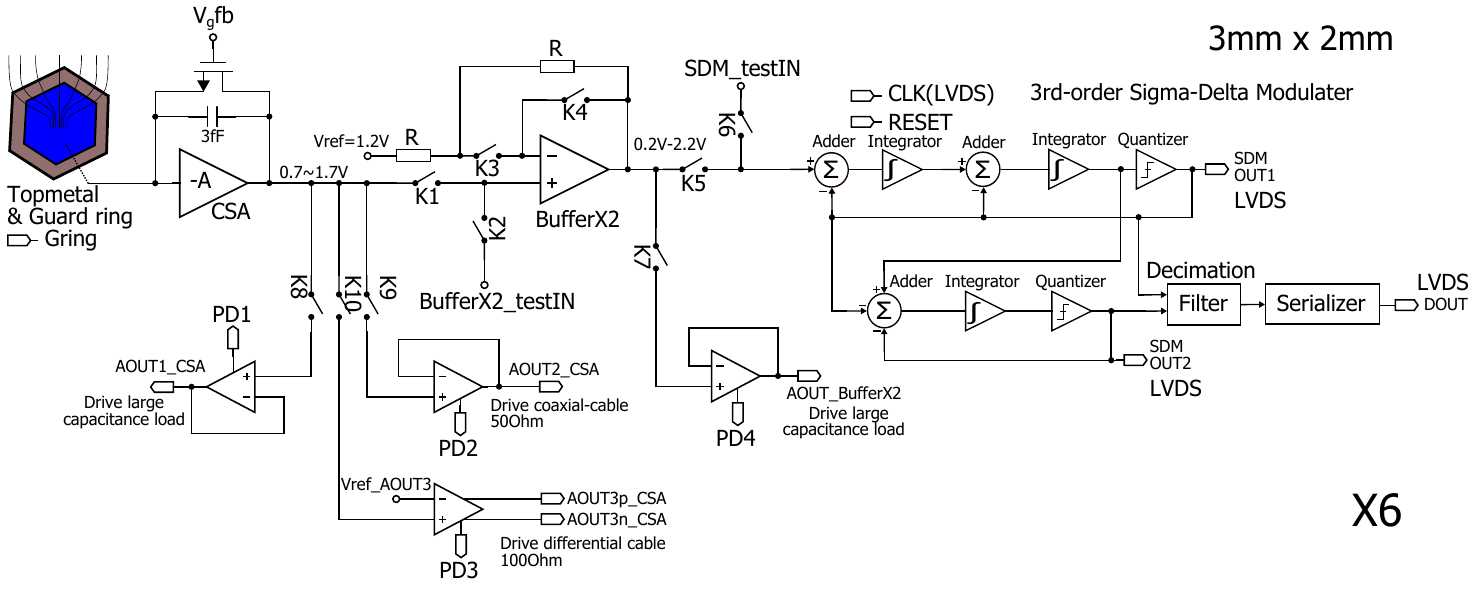}
  \end{minipage}\hspace{0.1em}%
  \begin{minipage}[t]{0.24\linewidth}
    (b)\\[7ex]
    \includegraphics[width=\linewidth]{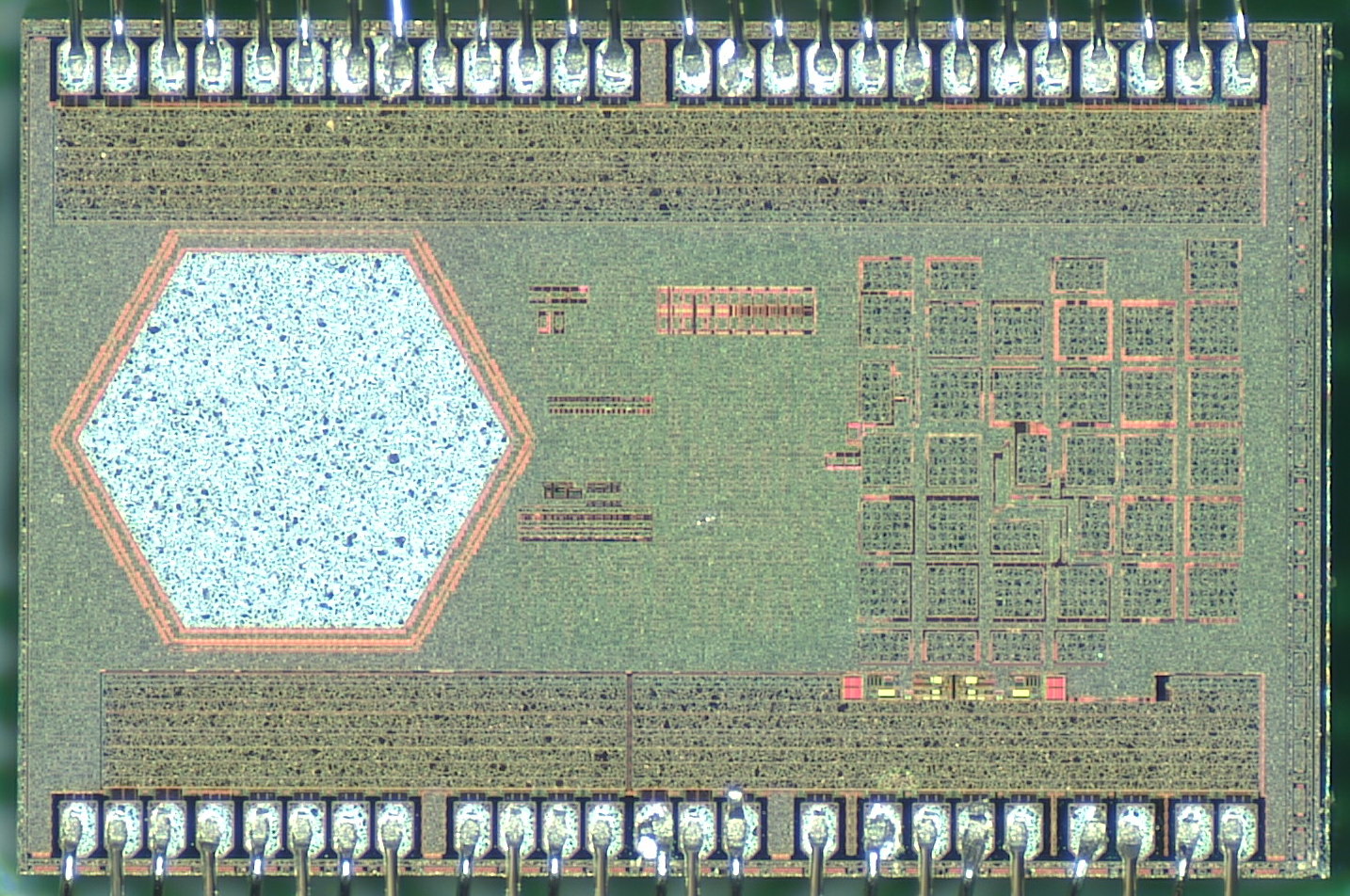}
  \end{minipage}
  \caption{\TMS sensor.  (a) A schematic overview of the sensor.  Each \TMS sensor consists of an electrode for direct charge collection, a Charge Sensitive pre-Amplifier (CSA), an internal $\times2$ buffer, and a 3rd-order Sigma-Delta Analog-to-Digital Converter (ADC).  Additional circuitry for diagnostics are present as well.  (b) Micrograph of a sensor.  The sensor is $3\times\SI{2}{mm^2}$ in size.  The \SI{1}{mm} sized hexagonal \TM electrode is located in the mid-left.}
  \label{fig:TMS}
\end{figure}

A top-level schematic view of the sensor is shown in Fig.~\ref{fig:TMS}(a).  The internal circuit of the CSA is in Fig.~\ref{fig:CSA}(a).  The output of the CSA is amplified by a factor of two then fed into a 3rd-order Sigma-Delta Modulator (SDM) that functions as an ADC.  The choice of an SDM was driven by the simplicity of its implementation.  SDMs allow the trade-off between the effective sampling rate and the signal-to-noise ratio at the data receiving end without any change to the circuitry.  Also, the SDM naturally outputs a bit-stream, eliminating the need of a data format, serializer, or transmission protocol.  These features are particularly appealing during the prototyping stage.

The sensor occupies a $3\times\SI{2}{mm^2}$ area, which satisfies the desired \SI{8}{mm} pitch size between sensors.  For debugging and prototyping purposes, many test and tuning points have been built into the design.  Nearly every module in the sensor could be independently validated.

\begin{figure}[!htb]
  \centering
  \begin{minipage}[t]{0.45\linewidth}
    (a)\\[3ex]
    \includegraphics[width=\linewidth]{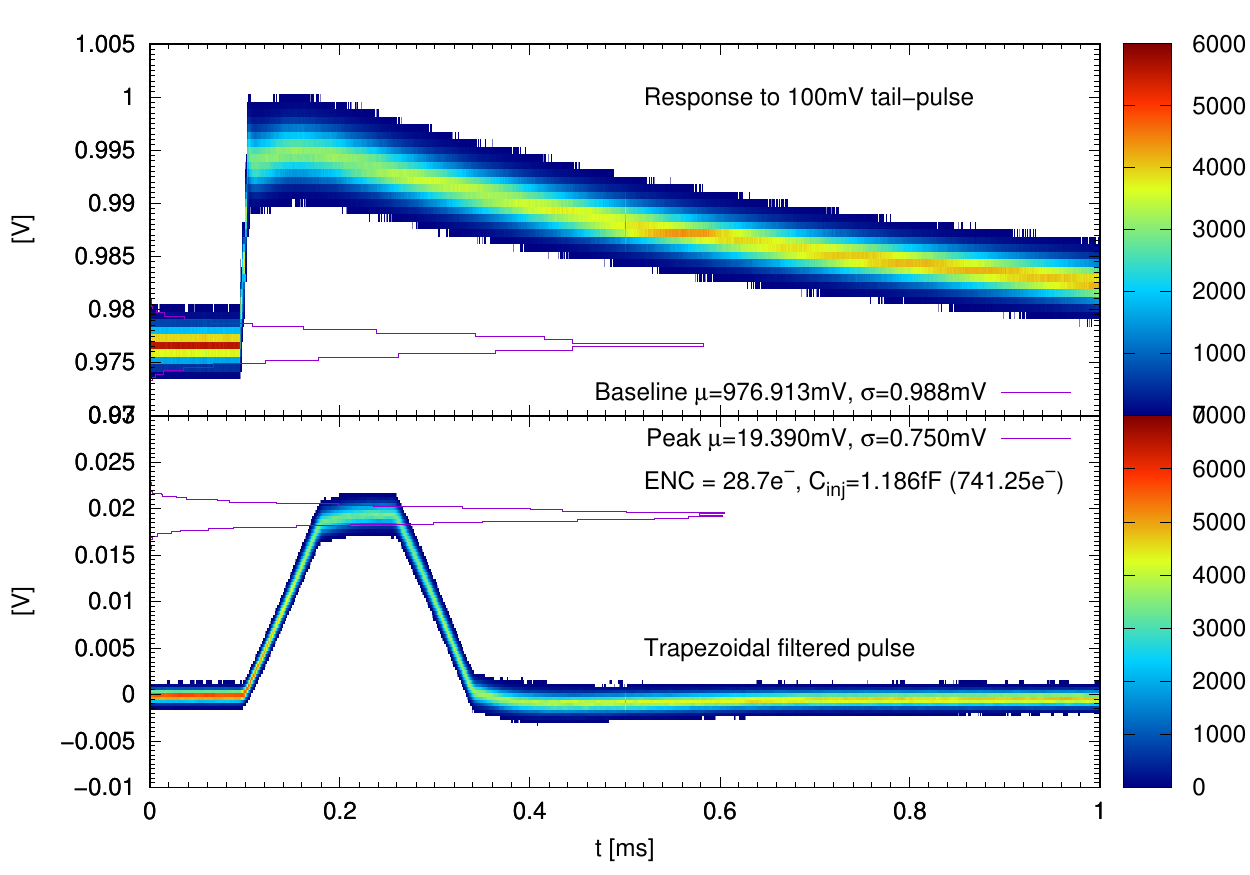}
  \end{minipage}%
  \begin{minipage}[t]{0.55\linewidth}
    (b)\\
    \includegraphics[width=\linewidth]{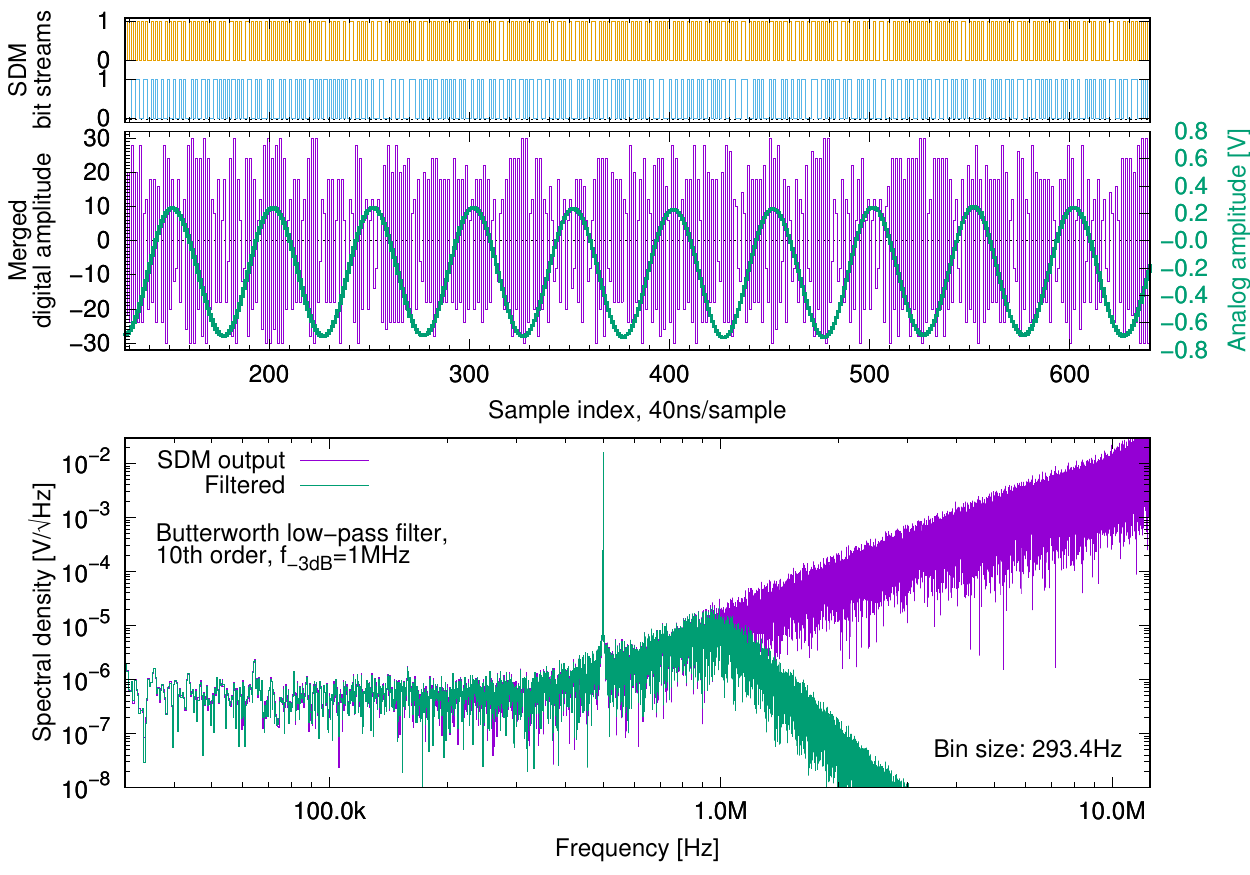}
  \end{minipage}
  \caption{(a) Charge Sensitive Amplifier response to test pulses.  Many test pulse responses are overlaid to form 2D histograms.  After applying a trapezoidal filter in software, an ENC$<\SI{30}{e^-}$ is achieved.  (b) Signals for Sigma-Delta Modulator in \TMS.  A \SI{500}{kHz} sine wave test signal with intentional DC offset is injected.  Top-panel shows the two bit-stream outputs from the SDM.  Mid-panel shows the combined digital data and recovered sine-wave signal.  Bottom-panel shows the frequency spectra behavior with and without the software filter used for signal recovery.}
  \label{fig:CSAnSDM}
\end{figure}

The first engineering run of \TMS has been completed.  Fig.~\ref{fig:TMS}(b) shows a picture of one sensor.  Using injected test pulses, we show that the electronic noise of CSA reached the desired $<\SI{30}{e^-}$ level, and the SDM ADC functions as expected as well (Fig.~\ref{fig:CSAnSDM}).  All data are streamed directly and continuously off the sensor.  A inter-sensor network will be investigated in a second version of the sensor.  We have also measured a small sample ($\sim\SI{10}{g}$ mass) of \TMS sensors in an underground low-background counting facility at SURF, which yielded upper-limits of radioactive contaminants consistent with those reported using other methods\cite{Leonard:2007uv,Albert2017}.  Since only up to \SI{1}{kg} CMOS material is going to be placed on a meter-sized plane, the radioactive background contribution from \TMS is negligible.

\subsection{19-sensor array}

\begin{figure}[!htb]
  \centering
  \begin{minipage}[t]{0.3\linewidth}
    (a)\\[25ex]
    \includegraphics[width=\linewidth]{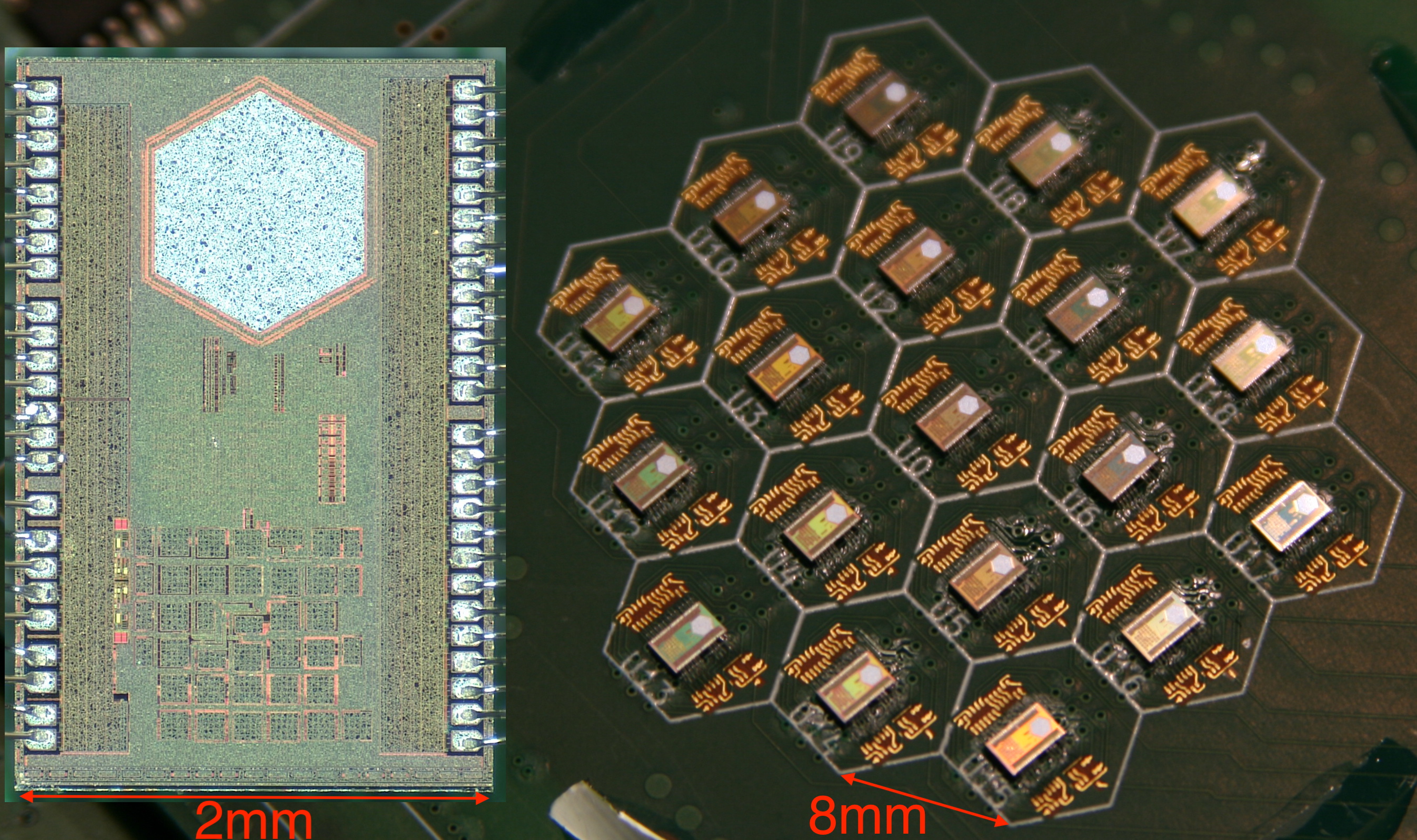}
  \end{minipage}%
  \begin{minipage}[t]{0.7\linewidth}
    (b)\\
    \includegraphics[width=\linewidth]{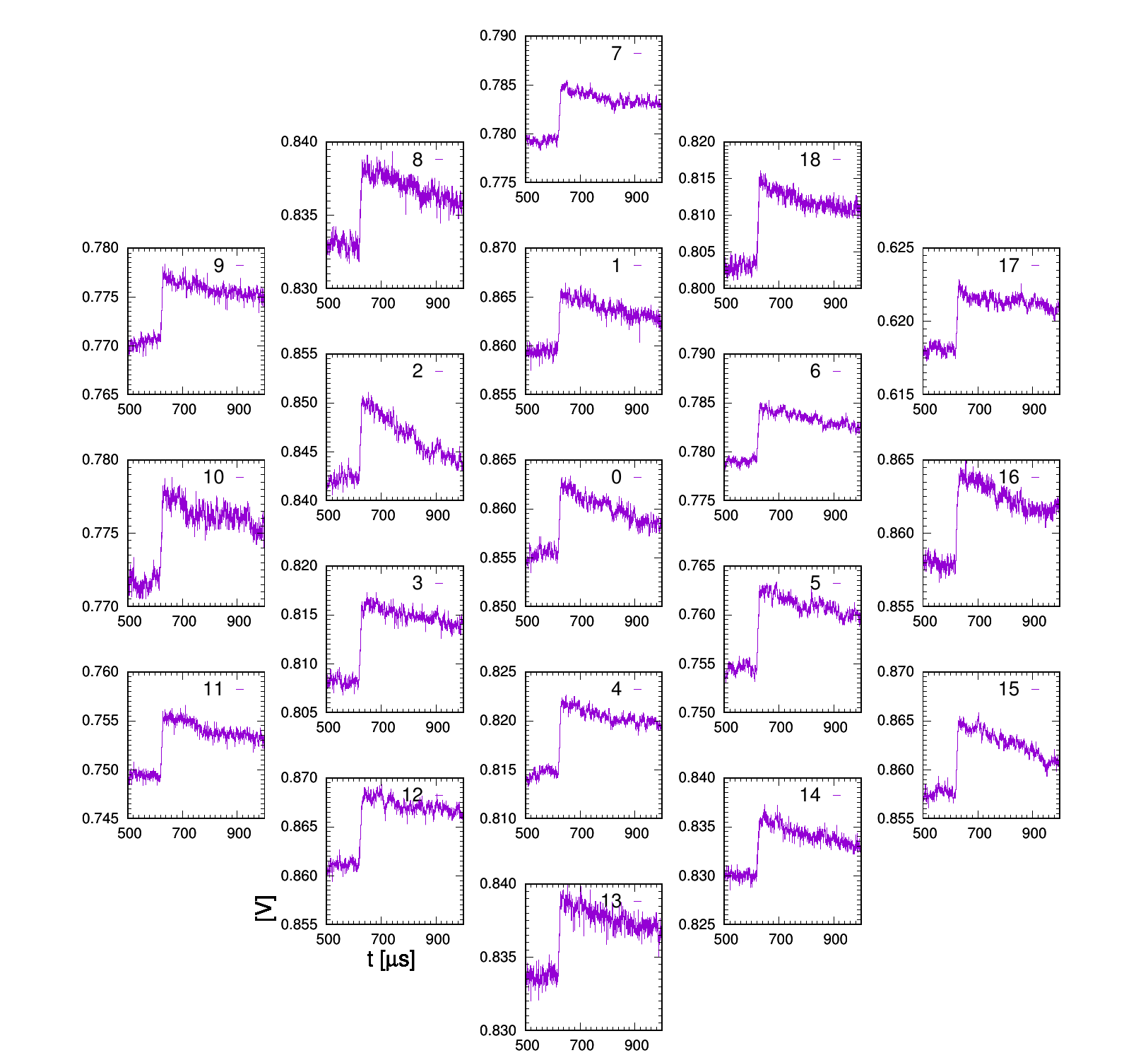}
  \end{minipage}
  \caption{(a) Picture of one sensor (enlarged) and a 19-sensor array.  (b) Response of every sensor in the array to a test pulse.}
\label{fig:TMS1mmX19}
\end{figure}

We assembled a small array consisting of 19 sensors with the optimal pitch size of \SI{8}{mm} (Fig.~\ref{fig:TMS1mmX19}(a)).  We developed a \texttt{python} script to generate the design of the PCB substrate.  It enables rapid change of array size.  We also established the chip selection capabilities using probe cards and probe stations to choose better performing sensors before assembly.  The selected sensors were placed and wire-bonded with automated machines.

FPGA-based supporting electronics that are required to collect and package data from all sensors have been implemented as well.  Waveforms are continuously captured by the entire system.  Fig.~\ref{fig:TMS1mmX19}(b) shows a snapshot of waveforms from all 19 sensors responding to a test pulse sent to them simultaneously.

\subsection{Future prospects and summary}

Through prior developments, we have demonstrated that \TM is capable of directly collect charge without avalanche gain, can integrate both analog and digital circuitry and have a low-enough noise to achieve $<\SI{1}{\percent}$ FWHM \znbb energy resolution.  A 19-sensor array with optimal pitch size of \SI{8}{mm} and data acquisition electronics have been implemented as well.

A small test TPC has been constructed to introduce drifting charge from gas to the sensor array to verify the performance in situ.  The results will be reported elsewhere.

The remaining challenges are centered around scaling up the system to a meter-sized plane.  We foresee the following path for this development.


A new \TMS version (v2) shall be developed to address the inter-sensor network issue described in Sec.~\ref{sec:sn} and to eliminate external component requirements described in Sec.~\ref{sec:pams}.  The \TMS v2 design will be completed by incorporating changes mandated by lessons learned from the characterization.

The large plane will be designed in a modular and staged fashion, and a few modules will be produced as a \emph{scaling prototype} to validate all the features and production procedures to \emph{scale-up} the size of the plane.  Data acquisition firmware and software will be adapted from the small-array system in the mean time.  A design correction to the \TMS v2 sensor and a second production is foreseen to incorporate necessary changes discovered from the \emph{scaling prototype}.

The \emph{scaling prototype} is expected to be about \SI{35}{cm} in diameter, which presents an opportunity for discovering potential problems at full-size plane deployment and characterization at medium-scale.  Since $\sim\si{MeV}$ $\beta$ travels many centimeters in high-pressure gas, a \SI{35}{cm} plane is the first chance to fully contain the energy in order to demonstrate energy resolution at \Qbb in gas.

The second \emph{scaling prototype} will be made using the \TMS v2 sensor with all corrections incorporated.  After gaining experience on the two \emph{scaling prototypes}, about 20 identical modules shall be produced, which will be tiled to realize the full-size plane.  Upon completion, the plane will be deployed into a high-pressure gaseous TPC chamber underground.

The above discussions were chiefly centered around the conditions for measuring electrons drifting in high-pressure gas without electron avalanche gain.  With proper tuning, mainly increasing the $RC$ constant of the CSA to many milliseconds, which can be done by writing a different bias value into the sensor digitally, the readout could detect ions drifting in gas directly.  No physical modifications are necessary.  This allows the very same system to use both Xe and \SeFs gases.  The swapping between different isotopes while keeping the identical detector construction is a unique advantage not offered by other techniques.

\acknowledgments Y.~Mei was supported by Laboratory Directed Research and Development (LDRD) funding from Berkeley Lab, provided by the Director, Office of Science, of the U.S.\ Department of Energy under Contract No.\ DE-AC02-05CH11231 for this work.  X.~Sun, from the Pixel Laboratory At CCNU (PLAC), was supported by the Young Talents Program of Central China Normal University (0900-31101200006).  We thank the \TM CMOS design team for their engineering contributions.



\bibliographystyle{JHEP}
\bibliography{refs}

\end{document}